\documentclass[prb,preprint,amsmath,amssymb,superscriptaddress,notitlepage]{revtex4-2}
\usepackage{epsfig}
\usepackage{graphicx}
\usepackage{color}
\usepackage{bm}
\usepackage{upgreek}
\usepackage{soul}

\begin{document}

\renewcommand{\figurename}{Fig.}
\renewcommand{\tablename}{Tab.}

\title{Altermagnetic lifting of Kramers spin degeneracy
}
\author{J.~Krempask\'y}
\thanks{These three authors contributed equally}
\affiliation{Photon Science Division, Paul Scherrer Institut, CH-5232 Villigen, Switzerland}

\author{L.~\v{S}mejkal}
\thanks{These three authors contributed equally}
\affiliation{Institut f\"ur Physik, Johannes Gutenberg Universit\"at Mainz, D-55099 Mainz, Germany}
\affiliation{Institute of Physics, Czech Academy of Sciences, Cukrovarnick\'{a} 10, 162 00 Praha 6 Czech Republic}

\author{S.~W.~D'Souza}
\thanks{These three authors contributed equally}
\affiliation{University of West Bohemia, New Technologies Research Center, Plzen 30100, Czech Republic}

\author{ M.~Hajlaoui}
\affiliation{Institute of Semiconductor and Solid State Physics,
Johannes Kepler University Linz, 4040 Linz, Austria}

\author{G.~Springholz}
\affiliation{Institute of Semiconductor and Solid State Physics,
Johannes Kepler University Linz, 4040 Linz, Austria}

\author{K.~Uhl\'i\v{r}ov\'a}
\affiliation{Faculty of Mathematics and Physics,
Charles University, Ke Karlovu 3, CZ-121 16, Prague 2, Czech Republic}

\author{F.~Alarab}
\affiliation{Photon Science Division, Paul Scherrer Institut, CH-5232 Villigen, Switzerland}

\author{P.~C.~Constantinou}
\affiliation{Photon Science Division, Paul Scherrer Institut, CH-5232 Villigen, Switzerland}

\author{V.~Strokov}
\affiliation{Photon Science Division, Paul Scherrer Institut, CH-5232 Villigen, Switzerland}

\author{ D.~ Usanov}
\affiliation{Photon Science Division, Paul Scherrer Institut, CH-5232 Villigen, Switzerland}

\author{ W.~R. Pudelko}
\affiliation{Photon Science Division, Paul Scherrer Institut, CH-5232 Villigen, Switzerland}

\author{R. Gonz\'{a}lez-Hern\'{a}ndez} 
\affiliation{Grupo de Investigaci\'{o}n en F\'{\i}sica Aplicada, Departamento de F\'{i}sica,Universidad del Norte, Barranquilla, Colombia}

\author{A.~Birk~Hellenes}
\affiliation{Institut f\"ur Physik, Johannes Gutenberg Universit\"at Mainz, D-55099 Mainz, Germany}

\author{Z.~Jansa}
\affiliation{University of West Bohemia, New Technologies Research Center, Plzen 30100, Czech Republic}

\author{H.~Reichlov\'a}
\affiliation{Institute of Physics, Czech Academy of Sciences, Cukrovarnick\'{a} 10, 162 00 Praha 6 Czech Republic} 

\author{Z.~\v{S}ob\'a\v{n}}
\affiliation{Institute of Physics, Czech Academy of Sciences, Cukrovarnick\'{a} 10, 162 00 Praha 6 Czech Republic}

\author{R.~D.~Gonzalez~Betancourt}
\affiliation{Institute of Physics, Czech Academy of Sciences, Cukrovarnick\'{a} 10, 162 00 Praha 6 Czech Republic}

\author{P.~Wadley}
\affiliation{School of Physics and Astronomy, University of Nottingham, Nottingham NG7 2RD, United Kingdom}

\author{J. Sinova}
\affiliation{Institut f\"ur Physik, Johannes Gutenberg Universit\"at Mainz, D-55099 Mainz, Germany}
\affiliation{Institute of Physics, Czech Academy of Sciences, Cukrovarnick\'{a} 10, 162 00 Praha 6 Czech Republic}

\author{D.~Kriegner}
\affiliation{Institute of Physics, Czech Academy of Sciences, Cukrovarnick\'{a} 10, 162 00 Praha 6 Czech Republic} 

\author{J. Min\'ar}
\thanks{Corresponding authors: juraj.krempasky@psi.ch, jminar@ntc.zcu.cz, jungw@fzu.cz}
\affiliation{University of West Bohemia, New Technologies Research Center, Plzen 30100, Czech Republic}

\author{ J.H.~Dil}
\affiliation{Institut de Physique, \'{E}cole Polytechnique F\'{e}d\'{e}rale de Lausanne, CH-1015 Lausanne, Switzerland}

\author{T.~Jungwirth}
\thanks{Corresponding authors: juraj.krempasky@psi.ch, jminar@ntc.zcu.cz, jungw@fzu.cz}
\affiliation{Institute of Physics, Czech Academy of Sciences, Cukrovarnick\'{a} 10, 162 00 Praha 6 Czech Republic}
\affiliation{School of Physics and Astronomy, University of Nottingham, Nottingham NG7 2RD, United Kingdom}

\maketitle

{\bf 
Lifted Kramers spin-degeneracy has been among the central topics of condensed-matter physics since the dawn of the band theory of solids \cite{Kramers1930,Wigner1932}. It underpins established practical applications as well as current frontier research, ranging from magnetic-memory technology \cite{Chappert2007,Ralph2008,Bader2010,Bhatti2017,Manchon2019} to topological quantum matter \cite{Nagaosa2010,Franz2013,Bradlyn2017,Smejkal2018,Zang2018,Tokura2019,Vergniory2019,Tokura2019,Xu2020,Elcoro2021,Smejkal2022AHEReview}.
Traditionally, lifted Kramers spin-degeneracy has been considered to originate from two possible internal symmetry-breaking mechanisms. The first one refers to time-reversal symmetry breaking by magnetization of ferromagnets, and tends to be strong due to the non-relativistic exchange-coupling origin \cite{Landau1984}. The second mechanism applies to crystals with broken inversion symmetry, and tends to be comparatively weaker as it originates from  the relativistic spin-orbit coupling \cite{Winkler2003,Armitage2018,Krempasky2016,DiSante2013}. A recent theory work based on spin-symmetry classification has identified an unconventional magnetic phase, dubbed altermagnetic \cite{Smejkal2021a,Smejkal2022a}, that allows for lifting the Kramers spin degeneracy without net magnetization and inversion-symmetry breaking. Here we provide the confirmation using photoemission spectroscopy and {\em ab initio} calculations. We identify two distinct unconventional mechanisms of lifted Kramers spin degeneracy generated by the altermagnetic phase of centrosymmetric MnTe with vanishing net magnetization \cite{Smejkal2021a,Smejkal2022a,Betancourt2021,Mazin2023}. Our observation of the altermagnetic lifting of the Kramers spin degeneracy can have broad consequences in magnetism. It motivates  exploration and exploitation of the unconventional nature of this magnetic phase in an extended family of materials, ranging from insulators and semiconductors to metals and superconductors \cite{Smejkal2021a,Smejkal2022a}, that have been either identified recently or perceived for many decades 
 as conventional antiferromagnets \cite{Smejkal2022a,Neel1971,Kunitomi1964}.
} 

A recently developed spin-symmetry classification focusing on collinear magnets and, within  the hierarchy of  interactions, on the strong non-relativistic exchange, has identified a third elementary type of magnetic phases in addition to the conventional ferromagnets and antiferromagnets \cite{Smejkal2021a,Smejkal2022a}. The exclusively distinct spin-symmetry characteristics of this emerging third, altermagnetic class are the opposite-spin sublattices connected by a real-space rotation transformation (proper or improper and symmorphic or non-symmorphic), but not connected by a translation or inversion  \cite{Smejkal2021a,Smejkal2022a}. In contrast, the conventional ferromagnetic (ferrimagnetic)  class has one spin lattice or opposite-spin sublattices not connected by any symmetry transformation, and the conventional antiferromagnetic class has opposite-spin sublattices connected by a real-space translation or inversion. For the case of inversion, the Kramers spin degeneracy of bands in these conventional antiferromagnets is protected even in the presence of the relativistic spin-orbit coupling \cite{Smejkal2017c}. For the translation connecting the  opposite-spin sublattices, lifting of the Kramers spin degeneracy in these antiferromagnets requires both spin-orbit coupling and inversion-symmetry breaking in the crystal, in analogy to ordinary non-magnetic systems.

The unconventional nature of altermagnets is that the rotation symmetry connecting the  opposite-spin sublattices protects an antiferromagnetic-like compensated magnetic order with a vanishing net magnetization while, simultanously, it enables a ferromagnetic-like   lifting of the Kramers spin degeneracy without breaking the crystal inversion symmetry and without additional symmetry breaking by the relativistic spin-orbit coupling \cite{Smejkal2021a,Smejkal2022a}. Here we will refer to this mechanism as "strong" altermagnetic lifting of the Kramers spin degeneracy. 

Apart from the signature antiferromagnetic-like vanishing magnetization and ferromagnetic-like strong spin-degeneracy lifting, whose presence have been traditionally considered as mutually exclusive in one physical system, altermagnets can host a range of novel phenomena that are unparalleled in either the conventional ferromagnets or antiferromagnets \cite{Smejkal2021a,Smejkal2022a}. Within the realm of (lifted) Kramers spin-degeneracy physics, a unique property associated with the alternating sign of the spin polarization in the altermagnet's Brillouin zone is the presence of an even number of spin-degenerate nodal surfaces crossing the zone-center ($\boldsymbol\Gamma$-point) in the non-relativistic band structure. In Fig.~1a we demonstrate that these spin degeneracies can be lifted by the relativistic spin-orbit coupling in altermagnets even without breaking the crystal inversion symmetry. We will refer to this mechanism as "weak" altermagnetic lifting of the Kramers spin degeneracy. A comparison of these unconventional weak and strong mechanisms of lifted Kramers spin degeneracy, enabled by altermagnetism, are illustrated in Figs.~1a,b.

\begin{figure}[h!]
\hspace*{0cm}\epsfig{width=1\columnwidth,angle=0,file=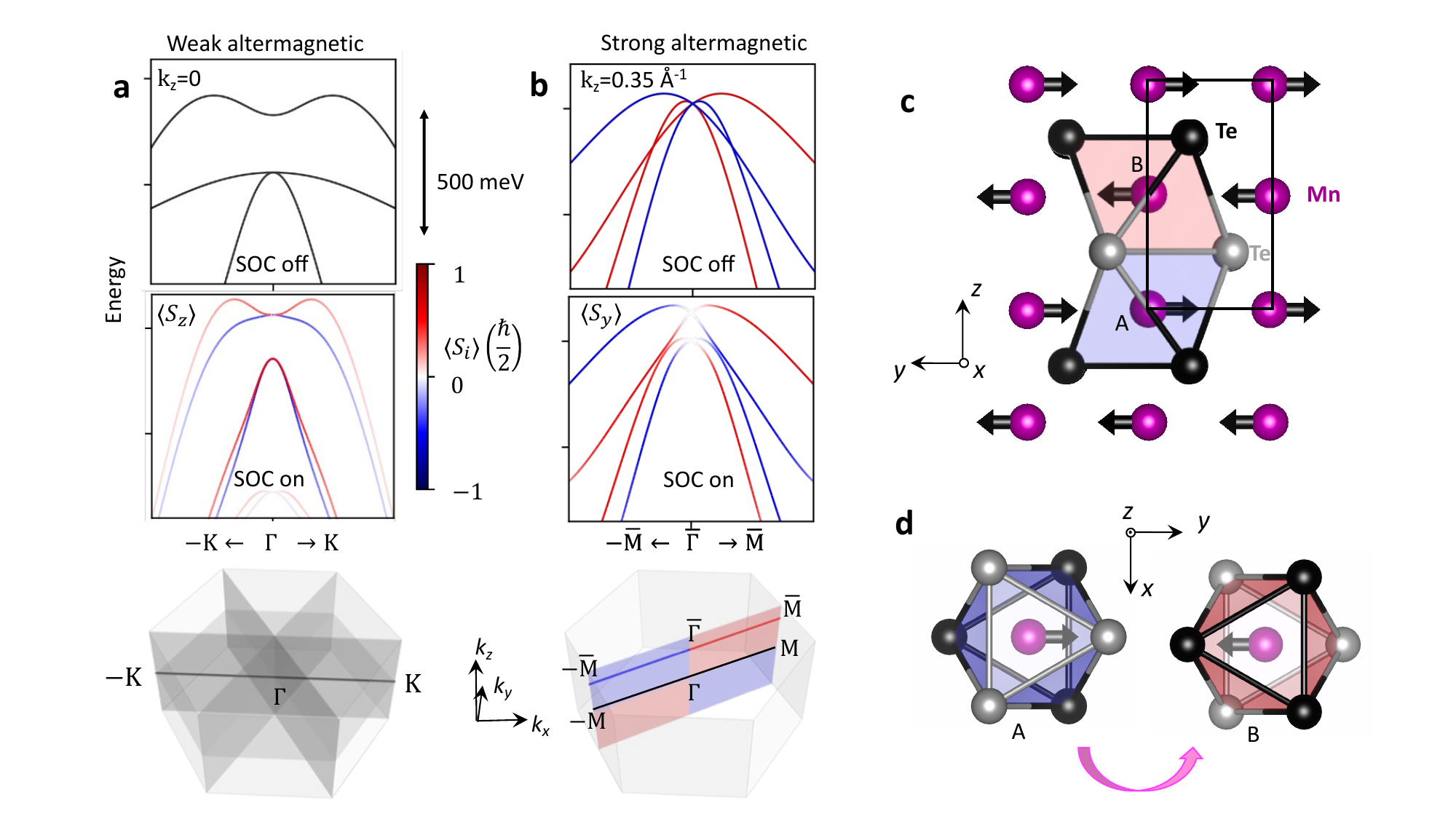}
\caption{\textbf{Illustration of weak and strong altermagnetic lifting of Kramers spin degeneracy.} 
\textbf{a,} Top and middle panels: {\em ab initio} band structure of MnTe at $k_z=0$ along the $\boldsymbol\Gamma - {\bf K}$ path for relativistic spin-orbit coupling turned off and on, resp. N\'eel vector is along the crystal $y$-axis (see panels c,d), corresponding to the $\boldsymbol\Gamma - {\bf M}$ axis (see bottom panel of {\bf b}). Red and blue colors correspond to opposite $z$-components of spin. Bottom panel: Schematics of the Brillouin zone with four spin-degenerate nodal planes in the electronic structure with spin-orbit coupling turned off.
\textbf{b,} Same as {\bf a} at $k_z=0.35$~\AA$^{-1}$ along the $\bar{\boldsymbol\Gamma} - \bar{{\bf M}}$ path illustrating the strong altermagnetic lifting of Kramers spin degeneracy. Red and blue colors correspond to opposite $y$-components of spin. Bottom panel highlights the $\bar{\boldsymbol\Gamma} - \bar{{\bf M}}$ path outside the four nodal planes (the red and blue colors highlight the alternating symmetry of the spin polarization in the plane).
\textbf{c,d} Schematic view of the crystal and magnetic structure of MnTe in the $y-z$ and $x-y$ plane, resp.
The red and blue shadings in \textbf{c,d} mark Te-octahedra around the Mn sites A and B with opposite spins which are related by spin rotation combined with six fold crystal rotation and half-unit cell translation along the $z$-axis.
}
\label{f1}
\end{figure}

Both the strong and the weak altermagnetic lifting of the Kramers spin degeneracy can enrich fields ranging from spintronics, ultrafast magnetism, magneto-electrics and magnonics, to topological matter, dissipationless quantum nanoelectronics and superconductivity  \cite{Smejkal2021a,Smejkal2022a}. For example, the strong altermagnetic lifting of the Kramers spin degeneracy has been theoretically shown to enable analogous spin-polarized currents to those used for reading and writing information in ferromagnetic memory devices while, simultaneously, removing the capacity and speed limitations imposed by a net magnetization  \cite{Smejkal2022a,Naka2019,Gonzalez-Hernandez2021,Naka2021,Ma2021,Smejkal2022,Smejkal2021a}. The weak altermagnetic lifting of the Kramers spin degeneracy has been linked to Berry-phase physics governing the dissipationless anomalous Hall currents while, again, removing the roadblocks associated with magnetization for realizing robust quantum-topological variants of these effects  \cite{Smejkal2022AHEReview,Smejkal2022a,Smejkal2020,Samanta2020,Naka2020,Hayami2021,Mazin2021,Betancourt2021,Naka2022}. Several of the predicted unconventional macroscopic time-reversal symmetry breaking responses accompanied by the vanishing magnetization  have been already experimentally confirmed in altermagnetic RuO$_2$ or MnTe \cite{Feng2022,Betancourt2021,Bose2022,Bai2021,Karube2022}. Here, using angle-resolved photoemission spectroscopy, we directly identify the weak and strong altermagnetic lifting of the Kramers spin degeneracy in the band structure of MnTe.

A schematic crystal structure of $\alpha$-MnTe  is shown in Fig.~1c,d. The two crystal sublattices A and B of Mn atoms, whose magnetic moments order antiparallel below the transition temperature of 310~K, are connected by a non-symmorphic six-fold screw-axis rotation, and are not connected by a translation or inversion \cite{Smejkal2021a,Betancourt2021}. The resulting non-relativistic electronic structure of this altermagnet is of the $g$-wave type \cite{Smejkal2021a} with three spin-degenerate nodal planes parallel to the $k_z$-axis 
and crossing $\boldsymbol\Gamma$  and {\bf K} points, and one additional spin degenerate nodal plane orthogonal to the $k_z$-axis and crossing the $\boldsymbol\Gamma$ point ($k_z=0$ nodal plane). These four nodal planes are highlighted in the bottom panel of Fig.~1a.

Angle-resolved photoemission spectroscopy (ARPES) measurements, shown in Fig.~2, were performed within the $k_z=0$ nodal plane along $k_x$ ($\boldsymbol\Gamma - {\bf K}$ path) and $k_y$ ($\boldsymbol\Gamma - {\bf M}$ path) using an X-ray photon energy of 667~eV. The experiments were performed on the soft X-ray ARPES beamline ADRESS at the Swiss Light Source synchrotron facility \cite{Strocov2010,Strocov2014}. Samples used in these measurements are thin MnTe films grown by molecular-beam epitaxy on a single-crystal InP(111)A substrate \cite{Kriegner2016,Betancourt2021}. We used a vacuum suitcase to transfer the thin-film samples from the growth to the soft X-ray ARPES chamber without breaking ultra-high-vacuum conditions (for details on the sample growth and characterization, and on the measurement techniques see Methods).

In Fig.~2a we show the measured raw data along the $k_x$-axis (bottom panel) and compare with one-step simulation of the photoemission process (top panel), using the Korringa-Kohn-Rostoker {\em ab-initio} approach that represents the electronic structure in terms of single-particle Green's functions \cite{Ebert2011a,Braun2018}. The intense spectral weight around -3.5~eV binding energy, indicated by a magenta dashed line in the experimental and theoretical panels of Fig.~2a, corresponds to a resonance due to Mn d-states. For a better visualization of the bulk electronic structure of MnTe, this spectral weight is filtered out in the experimental ARPES band maps shown in Figs.~2b,c. Refinements by the curvature mapping \cite{Zhang2011b} extracted from the area highlighted by a white-dashed rectangle are shown in  insets of top panels of Figs.~2b,c. These are compared to the corresponding relativistic {\em ab-initio} electronic structure calculations plotted in the bottom panels of Figs.~2b,c. The theoretical bands, with red and blue colors depicting opposite spin polarizations along the $z$-axis, show the weak altermagnetic lifting of the Kramers spin degeneracy within the $k_z=0$ nodal plane. The relativistic band-structure calculations were performed assuming the N\'eel vector along the in-plane $y$-axis (see Fig.~1c), consistent with earlier magnetic and magneto-transport measurements of the N\'eel-vector easy axis in epitaxial thin films of MnTe \cite{Kriegner2017,Betancourt2021}. Altermagnetism and spin-orbit coupling thus generate in this case an unconventional spin polarization of bands that is orthogonal to the direction of the magnetic-order vector.

\begin{figure}[h!]
\hspace*{0cm}\epsfig{width=1\columnwidth,angle=0,file=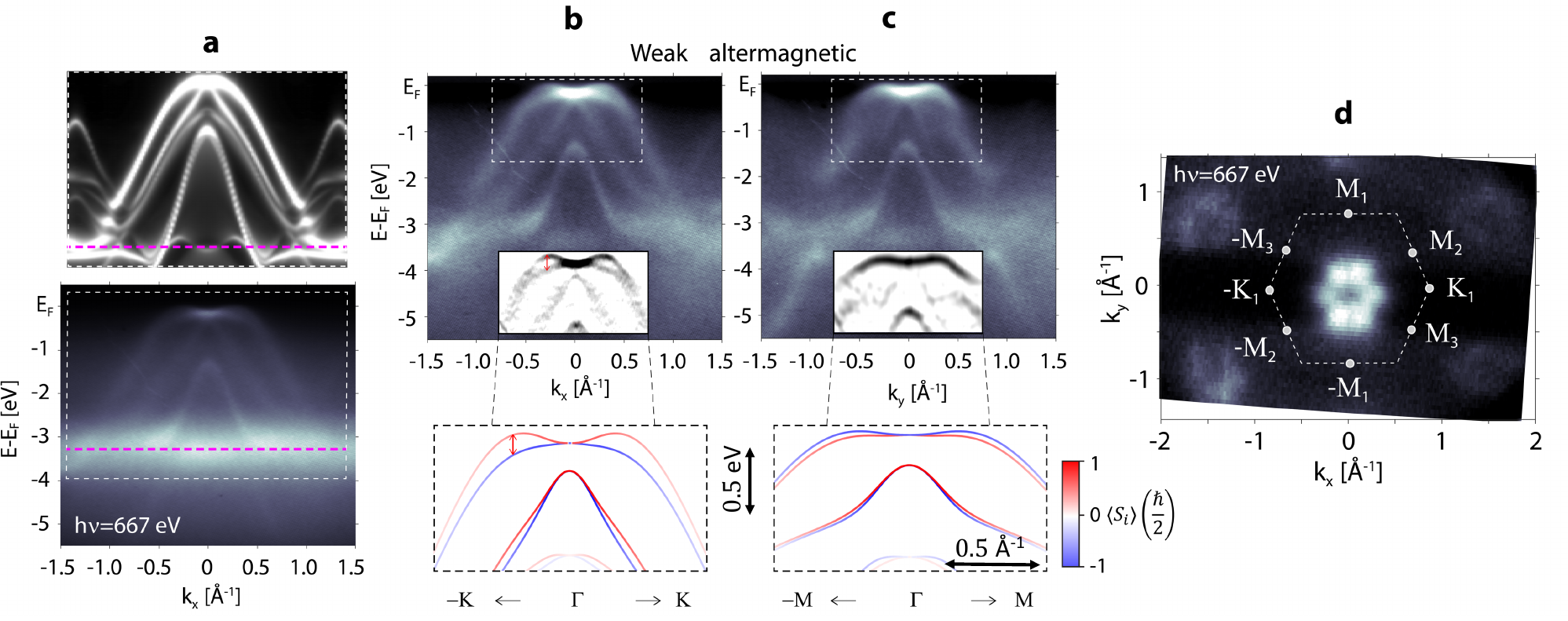}
\caption{\textbf{Weak altermagnetic lifting of Kramers spin degeneracy in the  nodal plane.} 
\textbf{a,} Bottom panel: Measured soft X-ray (667~eV) ARPES band map at $k_z=0$ along $k_x$ ($\boldsymbol\Gamma - {\bf K}$ path) on epitaxial thin-film MnTe. Top panel: Corresponding one-step ARPES simulation. Magenta dashed line highlights an intense spectral weight around -3.5~eV binding energy corresponding to a resonance of Mn d-states.
\textbf{b,} Measured ARPES band map along $k_x$ ($\boldsymbol\Gamma - {\bf K}_1$ path) after filtering out the intense spectral weight due to the Mn d-state  resonance.  Inset: Refinement of the measured data by curvature mapping. Bottom panel: {\em Ab initio} bands with red and blue colors corresponding to opposite $z$-components of spin. The N\'eel vector is aligned along the $\boldsymbol\Gamma - {\bf M}_2$ direction in the calculations.
\textbf{c,} Same as {\bf b} along $k_y$ ($\boldsymbol\Gamma - {\bf M}_1$ path). 
\textbf{d,} Constant-energy map  obtained by integrating the measured data over a 50~meV binding-energy interval from the top of the valence band.
}
\label{f2}
\end{figure}

The experimental ARPES band maps in Figs.~2b,c are fully consistent with the {\em ab-initio} band structures. This includes the overall band dispersions, as well as the significantly larger splitting of the top two bands along the $k_x$-axis ($\boldsymbol\Gamma - {\bf K}$ path, Fig.~2b) than along the $k_y$-axis ($\boldsymbol\Gamma - {\bf M}$ path, Fig.~2c). The splitting is highlighted in Fig.~2b by the red double-arrow in the experimental curvature map, and the two split bands have opposite spins in the corresponding {\em ab-initio} band structure. We again emphasize that this weak altermagnetic lifting of the Kramers spin degeneracy requires relativistic spin-orbit coupling and is unconventional as it is observed in the bulk band structure of an inversion-symmetric crystal. The extraordinary spin-splitting magnitude $\sim 100$~meV and the quadratic dispersion around the $\boldsymbol\Gamma$-point (see also Fig.~3c), consistently observed in experiment and theory, further highlight the unconventional  nature of this lifting of the Kramers spin degeneracy in altermagnetic MnTe. 

Fig.~2d shows a $k_z=0$ constant-energy map measured at the X-ray photon energy of 667~eV, obtained by integrating the measured data over a 50~meV interval of binding energies from the top of the valence band. The observed 6-fold symmetry indicates that within the probing area of this ARPES measurement, there is a comparable population of three N\'eel-vector easy axes, corresponding to the $\boldsymbol\Gamma - {\bf M}_{1-3}$ axes, that are crystallographically equivalent in the ideal hexagonal lattice of MnTe. Our observation of a multi-domain state is consistent with earlier magnetotransport measurements of the MnTe epilayers \cite{Kriegner2016,Betancourt2021}. 
We point out that domains with all these three N\'eel-vector easy axes exhibit larger spin splitting along $\boldsymbol\Gamma - {\bf K}_{1-3}$ paths than along $\boldsymbol\Gamma - {\bf M}_{1-3}$ paths, as shown in Extended Data Fig.~1. 
Therefore, even when the population of the three domains is comparable within the sample probing area  (X-ray spot position), a significantly larger splitting is expected for the $\boldsymbol\Gamma - {\bf K}_{1-3}$ paths than for the $\boldsymbol\Gamma - {\bf M}_{1-3}$ paths. This corroborates the excellent agreement between the experimentally observed and the calculated band splittings  in Fig.~2b,c.

The top-left panel of Fig.~3a shows the refinements by the curvature mapping corresponding to Fig. 2d. Together with the one-step ARPES simulation assuming an equal population of the three easy axes, shown in the top-right panel of Fig.~3a, it confirms the 6-fold symmetry of this constant-energy cut. In the series of panels in Fig.~3a, we then systematically explore the symmetry of the constant-energy maps measured and calculated at different  binding energies, indicated by symbols A-D in the band dispersion shown in Fig.~3c. Analogous set of measurements and calculations is shown in Fig.~3b for a different probing area on the sample (different X-ray spot position). While the maps in Fig.~3a show the 6-fold symmetry for all binding energies, the maps in Fig.~3b have a lower 2-fold symmetry at  energies near the top of the valence band (binding energies A-C). The 6-fold symmetry is observed in Fig.~3b only deeper in the valence band (binding energy D). The one-step ARPES simulations in Fig.~3b were performed assuming a single-domain state with the N\'eel vector along the easy axis corresponding to the $\boldsymbol\Gamma - {\bf M}_{1}$ axis. The agreement between experiment and theory for all the studied constant-energy maps confirms that in the probing area of the MnTe epilayer corresponding to Fig.~3b, there is a prevailing population of one of the three N\'eel-vector easy-axis domains ($\boldsymbol\Gamma - {\bf M}_{1}$ axis). A comparison to another probing area on the sample with an intermediate domain-population character between those  of Figs.~3a,b is shown in Extended Data Fig.~2. Note that, in Fig.~3b, the more prominent lowering of the symmetry from 6 to 2-fold near the top of the valence band correlates with the dominant contribution of $p$-orbitals of the heavy Te atoms, which significantly enhances the strength of spin-orbit coupling in this spectral range (see Extended Data Fig.~3). 

\begin{figure}[h!]
\hspace*{0cm}\epsfig{width=1\columnwidth,angle=0,file=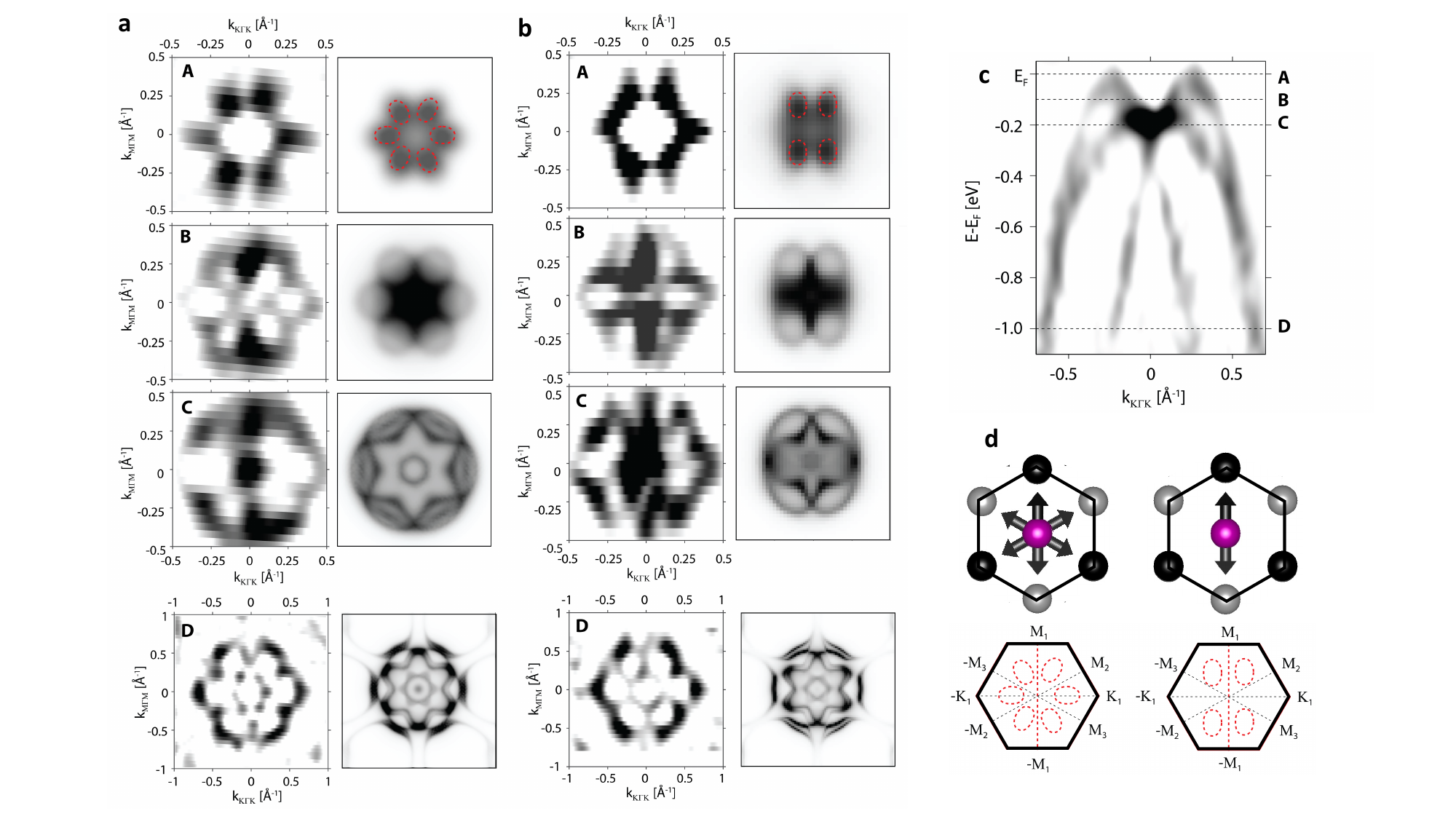}
\caption{\textbf{Constant-energy maps and N\'eel-vector easy-axis domains.} 
\textbf{a,} Left column: Refinements by the curvature mapping of measured constant-energy maps for binding energies A-D indicated in panel {\bf c}. Right column: Corresponding one-step ARPES simulations.
\textbf{b,} Same as {\bf a} for a different probing area on the sample (different X-ray spot position). Dashed red contours highlight the 6-fold (2-fold) symmetry in top right panels of {\bf a} ({\bf b}).
\textbf{c,} Refinement by the curvature mapping of the band map from the main experimental panel of Fig.~2a at $k_z=0$ along the $\boldsymbol\Gamma - {\bf K}$ path with the indicated binding energies A-D.
\textbf{d,} Schematics of the 6-fold symmetry of constant-energy maps (bottom-left)  for an equal (comparable) population of the three N\'eel-vector easy axes (top-left), and a lowered 2-fold symmetry of constant-energy maps (bottom-right) for one of the three easy-axes domains prevailing (top-right).
}
\label{f3}
\end{figure}

As explained in the introduction and illustrated in Fig.~1, the strong altermagnetic lifting of the Kramers spin degeneracy can be identified in the electronic structure only outside the four nodal planes that are spin-degenerate in the non-relativistic limit. In Fig.~4 we compare the measured and simulated ARPES data inside  and outside  the nodal planes. Soft X-ray ARPES band maps for $k_z=0.35$~\AA$^{-1}$
 (X-ray photon energy of 368~eV) along a path parallel to $\boldsymbol\Gamma - {\bf K}$, i.e., within one of the nodal planes, are shown in Figs.~4a,b. To highlight the finite $k_z$ value, we label the path as $\bar{\boldsymbol\Gamma} - \bar{{\bf K}}$. Data for the same $k_z$ value and a path $\bar{\boldsymbol\Gamma} - \bar{{\bf M}}$, i.e., outside  the nodal planes, are shown in Figs.~4c,d. In both experiment and theory, we observe a significantly larger band splitting in Figs.~4c,d (strong altermagnetic), reaching a half-eV scale, than Figs.~4a,b (weak altermagnetic) in the part of the spectrum labeled by B$_1$ and B$_2$. The spin-resolved one-step ARPES simulations of this part of the spectrum then suggest that a sizable spin-polarization signal should be detectable by spin-resolved ARPES (SARPES). This applies in particular to the $\bar{\boldsymbol\Gamma} - \bar{{\bf M}}$ path featuring the strong altermagnetic lifting of the spin degeneracy.

\begin{figure}[h!]
\hspace*{0cm}\epsfig{width=1\columnwidth,angle=0,file=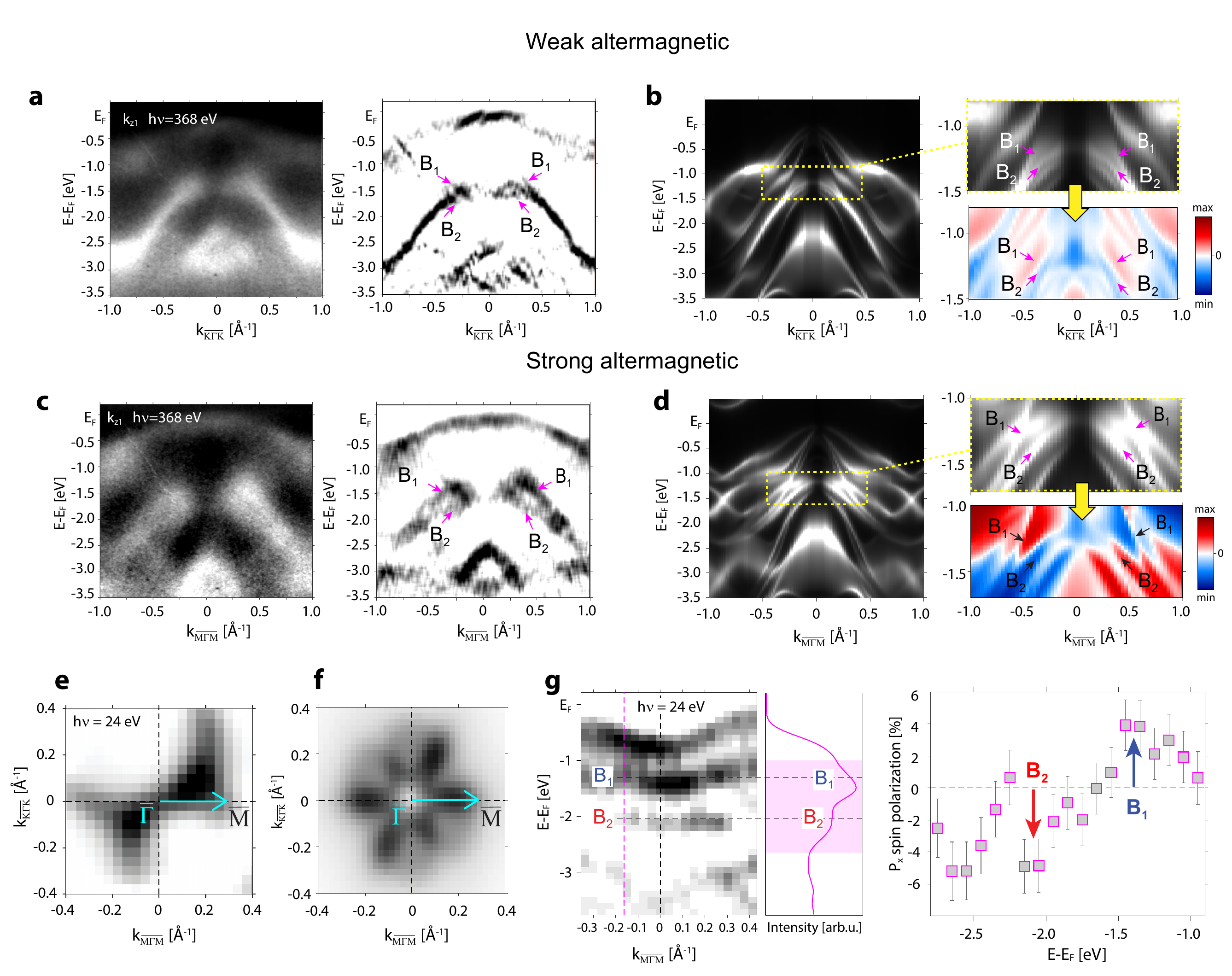}
\caption{\textbf{Weak and strong lifting of Kramers spin degeneracy at ${\bf k_z\neq 0}$.} 
\textbf{a,} Measured soft X-ray (368~eV) ARPES band map at $k_z=0.35$~\AA$^{-1}$ along the $\bar{\boldsymbol\Gamma} - \bar{{\bf K}}$ path (left: unrefined data, right: refined data). 
\textbf{b,} Corresponding one-step ARPES simulations. Red and blue colors show opposite $y$-components of spin. 
\textbf{c,d} Same as {\bf a,b} along the $\bar{\boldsymbol\Gamma} - \bar{{\bf M}}$ path.
\textbf{e,} Experimental UV (24~eV) ARPES constant-energy maps at $k_z=0.12$~\AA$^{-1}$ measured on bulk-crystal MnTe.
\textbf{f,} Corresponding one-step UV ARPES simulations.
\textbf{g,} Experimental UV ARPES band map along the $\bar{\boldsymbol\Gamma} - \bar{{\bf M}}$ path (left), corresponding total-intensity energy-distribution curve (middle) and SARPES (right). The spin polarization is detected along  the $\bar{\boldsymbol\Gamma} - \bar{{\bf M}}$ axis.
In all theoretical panels, the considered N\'eel vector and the spin-polarization projection are along an axis corresponding to the $\bar{\boldsymbol\Gamma} - \bar{{\bf M}}_1$ axis (also highlighted by the cyan arrow in panels e,f), and the considered paths are $\bar{\boldsymbol\Gamma} - \bar{{\bf K}}_1$ and $\bar{\boldsymbol\Gamma} - \bar{{\bf M}}_1$.
}
\label{f4}
\end{figure}

The spin-resolved measurements  were performed on the UV SARPES end-station COPHEE at the Swiss Light Source \cite{Hoesch2002}. Since the COPHEE sample holder was not compatible with the Omicron plate used  in the vacuum suitcase, and the system  did not allow for decapping the MnTe surface, we performed the SARPES measurements on {\em in situ} cleaved bulk-crystal samples. The bulk single crystals of MnTe were grown using the self-flux method. Their structural quality was confirmed by X-ray diffraction. Magnetization measurements by a superconducting quantum interference device verified the compensated magnetic ordering with the N\'eel temperature at 310~K and the N\'eel vector in the $z$-plane, consistent with earlier reports on bulk crystals \cite{ Kunitomi1964},  and also consistent with the magnetic characteristics of our thin MnTe films. (For more details on the characterization of the bulk MnTe crystals see Methods and Supplementary information.)

The consistency between the electronic structures of the MnTe thin-film and bulk-crystal samples is illustrated on soft X-ray ARPES band maps shown in Extended Data Figs.~3a,b, accompanied by the corresponding one-step ARPES simulations in Extended Data Figs.~3c. (Note that in both the experimental and theoretical band maps we consistently observe that the photoemission final state effects almost completely suppress the band mapping for $\boldsymbol\Gamma_2$ and $\boldsymbol\Gamma_4$.) The colored stripes in  experimental panels of Extended Data Fig.~3 highlight resonances due to Te-states (purple) and Mn-states (yellow-green), also observed in the simulations. In Extended Data Fig.~3d,e we accompany the ARPES data by plotting corresponding atomic-orbital projections of {\em ab initio}  bands and the density of states, consistently showing that the Te $p$-orbitals dominate the top of the valence band, while the spectral weight of Mn $d$-orbitals becomes significant below -3~eV. 

UV ARPES measurements  of the MnTe  bulk crystal at photon energy of 24~eV and the corresponding simulation of the constant energy map for non-zero $k_z$ ($k_z=0.12$~\AA$^{-1}$) are shown in Figs.~4e,f. The lowered 2-fold symmetry confirms the prevailing population of one of the three easy-axis domains in the probing area of the measured  bulk-crystal sample.  In the UV ARPES  band map along the $\bar{\boldsymbol\Gamma} - \bar{{\bf M}}$ path, plotted in the left panel of Fig.~4g, we identify the analogous spectral features to those labelled as B$_1$ and B$_2$  in Figs.~4c,d whose expected spin polarization is due to the strong altermagnetic lifting of the Kramers spin degeneracy. The spin polarization is experimentally confirmed by the UV SARPES measurements in Fig.~4g. In the middle panel of Fig.~4g we plot the measured total-intensity energy-distribution curve (EDC), and  the corresponding SARPES signal is shown in the right panel of Fig.~4g. As expected, we observe the alternating  sign of the spin-polarization component along the N\'eel vector, consistent with the presence of the strong altermagnetic lifting of the Kramers spin degeneracy for the $\bar{\boldsymbol\Gamma} - \bar{{\bf M}}$ path.  

In conclusion, we have observed two types of unconventional lifting of Kramers spin degeneracy in altermagnetic MnTe. The weak altermagnetic mechanism  generates extraordinary relativistic spin-splitting magnitude $\sim 100$~meV and quadratic dispersion around the $\boldsymbol\Gamma$-point. The strong altermagnetic mechanism in the magnetically compensated and centrosymmetric MnTe reaches a remarkable half-eV scale. The experimental observations are in excellent agreement with {\em ab initio} calculations. The agreement between the spin-split band structure observed in ARPES and obtained from density-functional theory confirms the prediction \cite{Smejkal2022a,Smejkal2022} that altermagnetism can originate directly from crystal symmetries, without requiring strong electronic correlations. Specifically, it confirms that altermagnetism stems from the local crystal anisotropy that breaks translation and inversion symmetry but preserves a rotation symmetry connecting the opposite-spin sublattices. The crystal-symmetry basis makes altermagnetism one of the elementary phases of matter which, remarkably, has been omitted for nearly a century of the band theory of solids. Our results highlight the strength of the spin-group symmetry classification in unraveling new magnetic phases and in describing the hierarchy of energy scales that underpin their rich phenomenology and potential applications \cite{Smejkal2022a,Smejkal2022,Smejkal2022AHEReview}.

\section*{Methods} 
{\em Thin film growth.} MnTe epilayers of\,200 nm thickness were grown by molecular beam epitaxy on single crystalline In-terminated InP(111) substrates using elemental Mn and Te sources. The less than 1\% lattice mismatch results in single crystalline hexagonal MnTe growth with the $c$-axis ($z$-axis) perpendicular to the surface. Two-dimensional growth of $\alpha$-MnTe is achieved at substrate temperatures of 370{--}450\,$^{\circ}$C. Further details on growth and sample characterization can be found in Ref.~\cite{Kriegner2017}. For ARPES experiments the samples were transferred after growth into a UHV suitcase in which they were transported  to the ARPES station at the synchrotron without breaking UHV conditions. 

{\em Single crystal growth.} For the growth of bulk MnTe crystals, pure manganese (99.9998~\%) and tellurium (99.9999~\%) in the molar composition Mn$_{33}$Te$_{67}$ were placed in an alumina (99.95~\%) crucible and, together with a catch crucible filled with quartz wool, sealed in a fused silica tube under vacuum. The sample was first heated up to 1050~$^{\circ}$C and then cooled down to 760~$^{\circ}$C for four days. At 760~$^{\circ}$C, the sample was quickly put into a centrifuge where the crystals were separated from the remaining melt. The crystals were in forms of  flat plates with lateral dimensions of several millimeters and thicknesses of hundreds of micrometers. 

{\em Characterization.}
Single crystal X-ray diffraction measurements were performed with a Rigaku Smartlab with 9 kW Cu rotating anode, Ge two-bounce monochromator and Hypix detector. Powder diffraction measurements for the lattice parameter determination were performed using a Panalytical Empyrean with Cu-tube in Bragg-Brentano geometry. Powder diffraction simulations were performed using $xrayutilities$ tool for reciprocal space conversion of scattering data \cite{kriegner2013}.
Magnetometry measurements were performed in a Quantum Design SQUID magnetometer using reciprocating sample option for increased measurement sensitivity. Temperature-dependent susceptibility measurements were taken in magnetic field of 50~mT.
Note that for the X-ray diffraction and SQUID magnetometry investigations samples were cleaned in aqua regia to remove a parasitic MnTe$_2$ phase formed at the surface during the final phases of the growth.
No traces of this phase could be detected in the X-ray diffraction investigations of the single crystals after this cleaning procedure. After cleaving the MnTe platelets therefore also expose a pristine $\alpha$-MnTe (0001) surface. 

{\em Photoelectron spectroscopy.}
Angle-resolved photoemission spectroscopy (ARPES) was used for investigating the electronic structure of MnTe - including the Fermi surface, band structure, and one-electron spectral function $A(\omega,k)$ - which are resolved in electron momentum $k$ (see Ref.~\cite{Damascelli_2003} for more details). 
The extension of photon energies into the soft X-ray range (SX-ARPES) from a few hundred eV to approximately 2 keV enhances the probing depth of this technique, characterised by the photoelectron escape depth $\lambda$, by a factor of 3-5 compared to the conventional vacuum ultraviolet photon energies (\hbox{UV-ARPES}). This enables access to the intrinsic bulk properties, which is essential for three-dimensional (3D) materials like MnTe. The increase of $\lambda$ reduces the intrinsic broadening $\delta{k_z}$ of the out-of-plane momentum $k_z$, defined by the Heisenberg uncertainty principle as  $\delta{k_z}\approx{\lambda}^{-1}$\,\cite{Strocov_VSe_2012}. Combined with the free-electron dispersion of high-energy final states, the resulting precise definition of $k_z$ allows accurate determination of the 3D electronic structure. As in the case of MnTe, this advantage of SX-ARPES has been demonstrated, e.g., also on ferroelectric Rashba semiconductors \cite{JK_PRB}, transition-metal dichalcogenides \cite{Strocov_VSe_2012,Weber_PRB_2018}, high-fold chiral fermion systems \cite{Schroeter_2020}, etc.

The SX-ARPES experiments were conducted in the photon energy range 350-700\,eV at the SX-ARPES end-station \cite{Strocov_JSR_2014} of the ADRESS beamline at the Swiss Light Source, Paul Scherrer Institute, Switzerland \cite{Strocov_JSR_2010}. All presented data were acquired with $\pi$-polarized X-rays. The photoelectrons were detected using the PHOIBOS-150 analyzer with an angular resolution of $\mathrm{\approx{0.1^{\circ}}}$ and using a deflector mode without changing the sample angles. The combined (beamline and analyzer) energy resolution varied between 50 and 100\,meV in the above energy range. The experiments were performed in a vacuum of better than $\mathrm{1{\times}10^{-10}}$\,mBar and at a sample temperature of around 15\,K. The investigated MnTe thin film samples were transferred from the MBE in JKU Linz using a vacuum suitcase. In the presented data, the coherent spectral fraction was enhanced by subtracting the angle-integrated spectral intensity as seen in Fig.\,2a-c of main text. The constant energy-surface maps were integrated within a range of $\pm${50}\,meV. The conversion of the measured photoelectron kinetic energies and emission angles to binding energies and momenta was accomplished using the kinematic formulas which account for the photon momentum \cite{Strocov_JSR_2014}. 

The spin-resolved ARPES (SARPES) measurements were conducted at 24\,eV at COPHEE experimental station at the Swiss Light Source SIS beamline \cite{Hoesch:2002, Dil_2009} on \textit{in situ} cleaved bulk single crystals at 21\,K. Combined with an angle-resolving photoelectron spectrometer it produces complete data sets consisting of photoemission intensities (Fig.\,4e), as well as spin polarization curves (Fig.\,4g) with the combined experimental resolution of $\approx$25\,meV and $\approx$100\,meV, respectively.

{\em Calculations.}
The experimental results were compared with \textit{ab initio} electronic structure calculations, performed for MnTe in P6$_\text{3}$/mmc (Space group:194)  symmetry using the lattice parameter as determined from the XRD measurements \cite{Kriegner2017}.

We calculated the electronic structure of MnTe in Fig.~1 and 2 with the pseudo-potential density functional theory code Vienna Ab initio Simulation Package (VASP) \cite{Kresse1996a}. %
 We used Perdew-Burge-Ernzerhof (PBE) \cite{Perdew1997}+SOC+U, a spherically invariant type of Hubbard parameter \cite{Betancourt2021} with a $8\times8\times5$ k-point grid, and a $520\,\mathrm{eV}$ energy cut-off.

The calculations in Fig.~2, 3 and 4 were carried out using spin-polarized fully relativistic Korringa-Kohn-Rostoker (SPRKKR) Green's function method in the atomic sphere approximation, within the rotationally invariant GGA+U scheme as implemented in the SPRKKR formalism \cite{Minar_RPP_2011,Minar_2011}. The screened on-site Coulomb interaction $U$ and  exchange interaction $J$ of Mn are set to 4.80\,eV and 0.80\,eV respectively.
The angular momentum expansion of the \textit{s,p,d,f} orbital wave-functions has been used for each atom on a 28$\times$28$\times$15 $k$-point grid. The energy convergence criterion has been set to $\mathrm{10^{-5}}$\,Ry. Lloyd's formula has been employed for accurate determination of the Fermi level \cite{Lloyd_1967,Lloyd_1972,Minar_2011}.

The  photoemission calculations for a semi-infinite surface of MnTe(001) with Mn atoms as the termination layer at the surface were performed within the one-step model of photoemission in the spin-density-matrix formulation as implemented in the SPRKKR package \cite{Braun_2018}.

\section*{Acknowledgement}
We acknowledge fruitful discussions with Karel V\'yborn\'y. This work was supported by the Czech Science Foundation project no. 19-28375X, Ministry of Education of the Czech Republic Grants LNSM-LNSpin, LM2018140,  and the Neuron Endowment Fund Grant. 
L.S. acknowledges support from JGU TopDyn initiative. L.S. and J.S. acknowledge the funding  by the Deutsche Forschungsgemeinschaft (DFG) Grant No. TRR 173 268565370 (project A03).
S.W.D and J.M thanks CEDAMNF project financed by the Ministry of
Education, Youth and Sports of Czech Republic, Project No. CZ.02.1.01/0.0/0.0/15 003/0000358 and  CZ.02.01.01/00/22 008/0004572, co-funded by the ERDF. WRP acknowledges support from the Swiss National Science Foundation (Projects No. 200021-185037). 
K.U. acknowledges the program of Czech Research Infrastructures (project no. LM2023065).
D.K. acknowledges the support from the Czech Academy of Sciences (project no. LQ100102201) and Czech Science Foundation (project no. 22-22000M)
M.H. and G.S. would like to acknowledge support by the Austrian Science Funds, Project P30960-N27 and I-4493-N.

\newpage

\bibliographystyle{naturemag}

\end{document}